\documentclass[conference]{IEEEtran}

\IEEEoverridecommandlockouts

\usepackage{graphicx}
\usepackage{amsmath}
\usepackage{amssymb}
\usepackage{color}

\usepackage{algorithm}
\usepackage{algpseudocode}
\algnewcommand\algorithmicinput{\textbf{Input:}}
\algnewcommand\INPUT{\item[\algorithmicinput]}
\algnewcommand\algorithmicoutput{\textbf{Output:}}
\algnewcommand\OUTPUT{\item[\algorithmicoutput]}

\usepackage{mathtools}
\DeclarePairedDelimiter{\ceil}{\lceil}{\rceil}
\DeclarePairedDelimiter{\floor}{\lfloor}{\rfloor}

\newcommand{\Fig}[1]{Fig.~\textup{\ref{#1}}}
\graphicspath{{figures/}}
\usepackage{subcaption}

\newtheorem{definition}{Definition}
\newtheorem{lemma}{Lemma}
\newtheorem{theorem}{Theorem}
\newtheorem{corollary}{Corollary}
\newtheorem{remark}{Remark}

\newcommand{\F}{\mathbb{F}}
\newcommand{\cC}{\mathcal{C}}
\newcommand{\cR}{\mathcal{R}}
\DeclareMathOperator{\Cl}{Cl} 

\begin{document}

\title{Bounds and Constructions of Codes with All-Symbol Locality and Availability \thanks{The research was carried out at the IITP RAS and supported by the Russian Science Foundation (project no. 14-50-00150).}}

\author{
  \IEEEauthorblockN{Stanislav Kruglik\IEEEauthorrefmark{1}\IEEEauthorrefmark{2}\IEEEauthorrefmark{3} and Alexey Frolov\IEEEauthorrefmark{1}\IEEEauthorrefmark{2}}
	
 \IEEEauthorblockA{\small \IEEEauthorrefmark{1} Skolkovo Institute of Science and Technology\\
    Moscow, Russia
    }
 \IEEEauthorblockA{\small \IEEEauthorrefmark{2} Institute for Information Transmission Problems\\
    Russian Academy of Sciences\\Moscow, Russia
    }
 \IEEEauthorblockA{\small \IEEEauthorrefmark{3} Moscow Institute of Physics and Technology \\
    Moscow, Russia
    }

  {stanislav.kruglik@skolkovotech.ru, al.frolov@skoltech.ru}
      
}


%


\maketitle

\begin{abstract}
 We investigate the distance properties of linear locally recoverable codes (LRC codes) with all-symbol locality and availability. New upper and lower bounds on the minimum distance of such codes are derived. The upper bound is based on the shortening method and improves existing shortening bounds. To reduce the gap in between upper and lower bounds we do not restrict the alphabet size and propose explicit constructions of codes with locality and availability via rank-metric codes. The first construction relies on expander graphs and is better in low rate region, the second construction utilizes LRC codes developed by Wang et al. as inner codes and better in high rate region.         

\end{abstract}

\section{Introduction}

A locally recoverable code (LRC) is a code over finite alphabet such that each symbol is a function of small number of other symbols that form a recovering set \cite{Gopp11,Gopp14,Papailiopoulos,Rawat,Yekhanin}. These codes are important due to their applications in distributed and cloud storage systems. LRC codes are well-investigated in the literature. The bounds on the rate and minimum code distance are given in \cite{Gopp11, Papailiopoulos} for the case of large alphabet size. The alphabet-dependent shortening bound (see \cite{Litsyn} for the method explanation) is proposed in \cite{Mazumdar}. Optimal code constructions are given in \cite{Silberstein} based on rank-metric codes (for large alphabet size, which is an exponential function of the code length) and in \cite{TamoBarg} based on Reed-Solomon codes (for small alphabet, which is a linear function of the code length).       

The natural generalization of an LRC code is an LRC code with availability (or multiple disjoint recovering sets).  Availability allows us to handle multiple simultaneous requests to erased symbol in parallel. This property is very important for hot data that is simultaneously requested by a large number of users. The case of LRC codes with availability is much less investigated. Bounds on parameters of such codes and constructions are given in \cite{Rawat, TamoBargFrolov, Parakash, Yaakobi}. Most of the papers focused on information-symbol locality and availability.  

We are interested in all-symbol locality and availability that is preferable in applications as it permits a uniform approach to system design. In this paper we continue the research started in \cite{TamoBargFrolov} and improve upper and lower bounds on the minimum distance of linear LRC codes with availability. To reduce the gap in between upper and lower bounds we do not restrict the alphabet size and propose explicit constructions of codes with locality  availability via rank-metric codes using the ideas from \cite{Silberstein}.

Our contribution is as follows. New upper and lower bounds on the minimum distance of LRC codes with availability are derived. The upper bound is based on the shortening method (developed in \cite{Litsyn}) and improves existing shortening bounds. We propose explicit constructions of LRC codes with availability via rank-metric codes. The first construction relies on expander graphs and is better in low rate region, the second construction utilizes codes with arbitrarily all-symbol locality and availability, high rate and small minimum distance developed in \cite{WangZhang} as inner codes and better in high rate region.           

\section{Preliminaries}

\subsection{Locally recoverable codes}

Let us denote by $\F_q$ a field with $q$ elements. Let $[n] = \{1, 2, \ldots, n\}$. The code $\cC \subset \F_q^n$ has locality $r$ if every symbol of the codeword $c\in \cC$ can be recovered from a subset of $r$ other symbols of $c$  \cite{Gopp11}. In other words, this means that, given $c\in \cC, i\in [n],$ there exists a subset of coordinates 
${\cR}_i\subset [n]\backslash i, |{\cR}_i|\le r$ such that the restriction of $\cC$ to the coordinates in ${\cR}_i$ enables one to find the value of $c_i.$ The subset ${\cR}_i$ is called a {\em recovering set} for the symbol $c_i$. 

Generalizing this concept, assume that every symbol of the code $\cC$ can be recovered from $t$ disjoint subsets of symbols of size $r$. More formally, denote by $\cC_I$ the restriction of the code $\cC$ to a subset of coordinates $I\subset [n]$.
Given $a\in \F_q$ define the set of codewords
   $
   \cC(i,a)=\{c\in \cC: c_i=a\},\; i\in[n].
   $

\begin{definition}
A code $\cC$  is said to have $t$ disjoint recovering sets if for every $i\in[n]$ there are $t$ pairwise disjoint subsets ${\cR}_{i}^1,\dots,{\cR}_{i}^t\subset [n]\backslash i$ such that for all $j=1,\dots,t$ and every pair of symbols $a,a'\in F_q, a\ne a'$
  \begin{equation*}\label{eq:rc}
  \cC(i,a)_{{\cR}_{i}^j}\cap \cC(i,a')_{{\cR}_{i}^j}=\emptyset.
  \end{equation*}
\end{definition}
 
In what follows we refer these codes as $(r,t)$-LRC codes. We briefly list the existing results below. The first bound for $(r,t)$-LRC codes was given in \cite{Wang2, Rawat2}
 \[
 d \leq n-k+2-\ceil*{\frac{t(k-1)+1}{t(r-1)+1}}.
 \]
  
An improvement of this bound was obtained in \cite{TamoBargFrolov}
\begin{equation*}\label{Barg}
  d \leq n - \sum_{i=0}^t \floor*{\frac{k-1}{r^i}}.
\end{equation*}
 
 An alphabet-dependent bound was probosed in \cite{Yaakobi} and has form 
  \begin{equation*}\label{Yaakobi}
  d \leq \min\limits_{\begin{subarray}{c}
 1\leq x \leq \ceil*{\frac{k-1}{(r-1)t+1}};{}{} 1 \leq y_j \leq t;{}{} j \in [x] \\ A<k;{}{}x,y_j\in Z^{+}
 \end{subarray}} d_{l-opt}^{q}[n-B,k-A] , 
  \end{equation*}
 where $A=\sum_{j=1}^x (r-1)y_j+x$, $B=\sum_{j=1}^x ry_j+x$ and $d_{l-opt}^{q}$ denote the
largest possible minimum distance of a code over $\F_q$.
 


The bound on the rate of $(r,t)$-LRC codes was given in \cite{TamoBargFrolov}
\begin{equation}\label{eq::rate}
\frac{k}{n} \leq R^*(r,t) = \prod_{i=1}^{t} \frac{1}{1+\frac{1}{ir}}. 
\end{equation}

This bound was improved in \cite{Parakash} for $t=2$.
 
In \cite{WangZhang} a recursive construction of binary $(r,t)$-LRC codes was proposed. The parameters of these codes are as follows: $n=\binom{r+t}{t}$, $R=\frac{r}{r+t}$ and $d = t+1$. We refer these codes as WZL codes and use them as inner codes in our constructions. We note, that in case of $t=2$ the construction of WZL codes coincides with the construction from \cite{TamoBargFrolov}. 

\subsection{Rank-metric codes}

\begin{definition}
A linearized polynomial $f(x)$ over $\F_{q^m}$ of $q$-degree $\ell$ can be presented as follows
\[
f(x) = \sum\limits_{i = 0}^{\ell} a_i x^{[i]}, 
\]
where $a_i \in \F_{q^m}$, $i=0, \ldots, \ell$, $a_\ell \ne 0$ and $x^{[i]} = x^{q^i}$.
\end{definition}

We now explain how to construct a codeword of $[n_G, k_G]$ Gabidulin code \cite{Gabidulin}. Let us choose an arbitrary linearized polynomial $f(x)$ over $\F_{q^m}$, such that  $q$-degree is less or equal to $k-1$. This polynomial includes $k$ information symbols as coefficients. Then
\[
c_G = ( f(\alpha_1), f(\alpha_2), \ldots, f(\alpha_{n_G})),
\]
where the elements $\alpha_1, \alpha_2, \ldots, \alpha_{n_G} \in \F_{q^m}$ and linearly independent as vectors (of length $m$) over $\F_{q}$. In what follows we assume $m \geq n$, we need this condition for $n$ linearly independent vectors to exist.

Note, that the following property of linearized polynomials holds
\begin{equation}\label{eq::linearized}
f(a\beta + b\gamma) = a f(\beta) + b f(\gamma),
\end{equation}
where $a, b \in \F_q$ and $\beta, \gamma \in \F_{q^m}$.

\subsection{Expander graphs}

Let us consider a biregular bipartite graph $G=(V \cup C, E)$ such that $|V| = n$ and $\deg v = t$ for $v \in V$, $|C| = \frac{nt}{r+1}$ and $\deg c=r+1$ for $c \in C$.

\begin{definition}
$G$ is an $(t, r+1, \alpha, t\gamma)$-expander if for any subset $V' \subset V$
\[
|V'| \leq \alpha n \Rightarrow |\Gamma(V')| > t \gamma |V'|,
\]
where $\Gamma(V') \subseteq C$ is the set of vertexes connected to the set $V'$.
\end{definition}

\begin{figure}[t]
\centering
\includegraphics[width=0.4\textwidth]{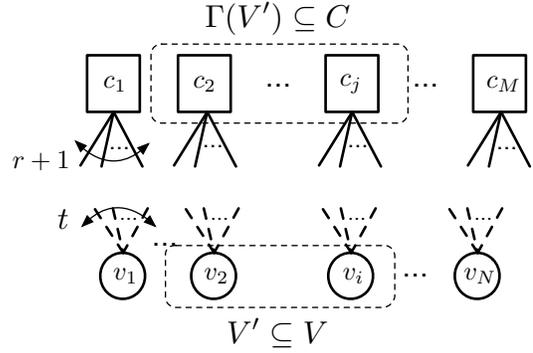}
\caption{Biregular expander graph}
\label{fig::expander}
\end{figure}

The definition is illustarted in \Fig{fig::expander}.

The usual way to check the expansion properties of a graph is to examine its second-largest eigenvalue (see e.g. \textup{\cite{A}}). Unfortunately, the explicit constructions of expander graphs with expansion greater than $t/2$ are not known \textup(Kahale \textup{\cite{Kahale}} even shows that eigenvalue separation cannot certify greater expansion\textup). Thus, in what follows we rely on the expansion properties of random expander graphs. The following asymptotic result, due to \cite{bur01}, is cited here in the form given in \cite[p.~431]{RU2008}.

\begin{lemma} %
\label{bursh}
Let $G$ be a graph chosen uniformly from the ensemble of $(t,r+1)$-regular bipartite graphs and let $n\to\infty.$ For a given $\gamma\in [\frac 1{r+1},1-\frac{1}{t})$ let $\delta$ 
be the positive solution of the equation 
  \begin{eqnarray*}
 \frac{t-1}{t} h(\delta)&-&\frac{1}{r+1} h(\delta\gamma(r+1))\nonumber\\ 
 &-&\delta\gamma(r+1) h\Big(\frac{1}{\gamma(r+1)}\Big)=0.
\label{eq:bursh}
 \end{eqnarray*} 
Then for $0<\delta'<\delta$ and $\beta=t(1-\gamma)-1$
\begin{equation*}
\Pr(\{G \text{ is an} (t,r+1,\delta', t\gamma) \text{ expander}\})\geq 1-O(n^{-\beta}).
\label{eq:bursh1}
\end{equation*}
\end{lemma}

\section{Upper bound on the minimum distance}

Shortening is a well known and widely used technique in coding theory. The idea is to remove (fix) some coordinates of the original code and use the new code to obtain bounds for the original code. 

Denote by $\Cl(I)$ the set of all coordinates such that for every $c \in \cC$ the values $c_i, i \in \Cl(I)$ can be found from the values of $c_I$. We will call the subset $\Cl(I)\supset I$ the {\em closure} of $I$ in $[n]$.

Let us introduce some notations. By $k^*(q, n, d)$ we denote an upper bound on the dimension of any linear code. By $d^*(q, n, d)$ we denote an upper bound on the distance of any linear code. The shortening bound can be formulated as follows.

\begin{theorem}[Shortening bound]\label{theorem::short}
Assume we are given an $[n, k, d]$ linear code $\cC$ over $\F_q$. The following inequalities hold for the parameters of the code
\[
k \leq \min\limits_{I: |\Cl(I)| \leq n-d} \left\{ |I| + k^*(q, n-|\Cl(I)|, d) \right\}
\]
and
\[
d \leq \min\limits_{I: |I| < k} \left\{ d^*(q, n-|\Cl(I)|, k-|I|) \right\}.
\]
\end{theorem}

\begin{remark}
We note, that the theorem~\ref{theorem::short} also valid in non-linear case. In this case by $k$ we mean $\log_q \cC$.
\end{remark}

Now we explain how the special structure of LRC code enables us to apply the shortening technique most efficiently. The result is formulates in the following theorem.

\begin{theorem}
Let $t \geq 2$. Assume we are given an $[n, k, d]$ linear $(r, t)$-LRC code $\cC$ over $\F_q$, then the following inequalities hold for the parameters of the code
\[
k \leq \min\limits_{sr+1 \leq n-d}(1+(r-1)s+k^*(q, n-1-sr, d)
\]
and
\[
d \leq \min\limits_{1+(r-1)s < k} d^*(n-1-sr, k - 1-(r-1)s).
\]
\end{theorem}
\begin{IEEEproof}

Within the proof we construct a set of coordinates $I$, $|I| = 1+(r-1)s$ with such a property
\[
|\Cl(I)| \geq 1+rs.
\]

Let us denote the code dual to $\cC$ by $\cC^\bot$. By $\cC^\bot_{r+1}$ we denote the set of codewords of dual code with the weight\footnote{Here and in what follows by weight we mean the Hamming weight, i.e. a number of non-zero elements in a vector.} less or equal to $r+1$ (local checks), i.e.
\[
\cC^\bot_{r+1} = \left\{ h \in \cC^\bot: \mathop{wt}(h) \leq r+1 \right\}.
\]
In what follows we work only with the set of all local checks $\cC^\bot_{r+1}$.

To construct the required set of coordinates $I$,$|I| = 1+(r-1)s$, we apply the Algorithm~\ref{alg::I} with input parameters $\cC^\bot_{r+1}$ and $s$. Let us explain the algorithm in more detail. At each step the algorithm adds a new local check (from the set $\cC^\bot_{r+1}$) to the set $X$ until $s$ linearly independent local checks are added. By $J$ we denote the set of covered positions. The algorithm chooses a local check with the largest intersection with $J$ (line~\ref{line::int}). Two cases are possible:
\begin{enumerate}
\item there exists a local check, which intersects with $J$.
\item there is no new local check, which intersects with $J$.
\end{enumerate}  

In the first case we need to check linear dependency (to proper calculate the number of check symbols) and add the local check to $X$. The second case is more interesting. This condition means, that the elements of $X$ form an $(r, t)$-LRC code of smaller length. Indeed the absence of new local checks, which intersects with at least one element of $X$ means that each position is covered either $t$ times or not covered at all. We store the number of recovery sets, that from an $(r, t)$-LRC code of smaller length in the variable $j$ and the number of check symbols of this code in the variable $s_1$.

It is clear, that the algorithm constructs the set $I$, such that $|\Cl(I) \backslash I| = s$. The only thing to check is that $|I|$ cannot be bigger then $1+(r-1)s$. We know, that the first $j$ elements of $X$ form an $(r, t)$-LRC code with $s_1$ check symbols. The worst case for the rest $s-s_1$ elements of $X$ is to intersect in exactly one position, so
\[
|I| \leq \frac{s_1}{1-R^*(r,t)}-s_1 + 1 + (s-s_1)(r-1) \leq 1 + s(r-1)
\]
as for $t \geq 2$ (see (\ref{eq::rate}))
\[
R^*(r,t) \leq \frac{r-1}{r}.
\]

\end{IEEEproof}

\begin{corollary}
If we substitute the Singleton bound for $d^*(q, n, d)$ function we obtain
\[
d \leq n-(k-1)- \floor*{ \frac{k-2}{r-1} }.
\]
\end{corollary}

\begin{corollary}
The asymptotic form of the new upper bound is as follows
\[
R \geq \frac{r-1}{r} (1-\delta) - o(1).
\]
\end{corollary}

\begin{algorithm}[t]
\caption{Construction of the set $I$}
\begin{algorithmic}[1]
\INPUT
\Statex $\cC^\bot_{r+1}$, $s$
\OUTPUT
\Statex $\mathbf{X}$, $I$, $s_1$, $j$

\State $H \gets \cC^\bot_{r+1}$
\State choose any $h \in H$
\State $J \gets \mathop{supp(\mathbf{h})}$, $X \gets \{h\}$, $H \gets H \backslash h$
\State $l \gets 1$ \Comment Number of added local checks
\State $i \gets 1$ \Comment Number of added linearly independent local checks
\State $j \gets 0$ 

\While {$i \leq s$}
	\State find the element $\mathbf{h} \in H$ with the largest $|J \cap \mathop{supp}(\mathbf{h})|$\label{line::int}     
    \If {$ |J \cap \mathop{supp}(\mathbf{h})| = 0$}
		\State $j \gets l$
		\State $s_1 \gets i$
		\State $i \gets i+1$
	\Else
		\If {$h \notin \mathop{span}\{X\}$}
			\State $i \gets i+1$
	    \EndIf
	    \State $J \gets J \cup \mathop{supp(\mathbf{h})}$, $X \gets X \cup \{h\}$, $H \gets H \backslash h$ 
	\EndIf
	\State $l \gets l+1$
\EndWhile
\State find $I$ from $X$ \Comment Note, that $J=\Cl(I)$
\If {$|I| < 1+(r-1)s$}
	\State add any $1+(r-1)s-|I|$ other coordinates
\EndIf
\end{algorithmic}
\label{alg::I}
\end{algorithm}

\section{Expander-based constructions}

In this section we show the existence of an $(r,t)$-LRC codes over a sufficiently large finite field $\F_{q^m}$ with large minimum distance. The proof relies on the existence of regular bipartite graphs with good expansion properties. We note, that the result here coincides with the result from \cite{TamoBargFrolov}. At the same time the construction is explicit and the proof is simpler.

Let $G=(V \cup C, E)$ be a bipartite graph with the following properties:
\begin{itemize}
\item $|V| = n$ and $\deg v = t$ for $v \in V$;
\item $|C| = \frac{nt}{r+1}$ and $\deg c=r+1$ for $c \in C$;
\item G is an $(t, r+1, \alpha, t \gamma)$-expander;
\item $\mathop{girth}(G) > 4$.
\end{itemize}

\begin{remark}
As shown in \cite{mckay04}, the probability that a random regular graph on $n$ vertexes has no cycles of length $4$ is bounded away from zero as $n\to\infty.$ This results together with Lemma~\ref{bursh} imply that there exist $(t,r+1,\delta, t\gamma)$ biregular bipartite expanding graphs with required properties.
\end{remark}

We now construct a matrix $\mathbf{H}_E = \left[ h_{j,i} \right]$, $1 \leq j \leq m$, $1 \leq i \leq n$, over $\F_q$. We associate the columns of $\mathbf{H}_E$ with the vertexes from $V$ and the rows of $\mathbf{H}_E$ with the vertexes of $C$. The element $h_{j,i}$ is non-zero if and only if the vertexes $v_i$ and $c_j$ are connected with an edge. We choose non-zero elements equiprobably and independently from the set $\F_q\backslash\{0\}$. By $\cC_E$ we denote a linear code of length $n$ over $\F_q$ determined by $\mathbf{H}_E$. The following inequality follows for the rate of the code
\[
R(\cC_E) \geq 1 - \frac{t}{r+1} - o(1),
\]
the equality takes place in case of full rank of $\mathbf{H}_E$.

Let us consider a code $\cC_G \diamondsuit \cC_E$ over $\F_{q^m}$, which is constructed in the following way. We first encode $k_G = k$ information symbols with $[k_G, n_G, d_G = n_G-k_G+1]$ Gabidulin code. Then we encode the resulting codeword of Gabidulin code with $[n = \frac{r+1}{r+1-t} n_G, n_G]$ code $\cC_E$. 
 
\begin{theorem}
Let us denote the relative minimum distance of the code $\cC_G \diamondsuit \cC_E$ by $\delta$. For sufficiently large $n$ and $q$ the following inequality holds for the rate $R$ of the code $\cC_G \diamondsuit \cC_E$
\[
R \geq 1 - \frac{t}{r+1} - \max\left\{ \delta (1 - t \gamma), 0 \right\} - o(1),
\]
where $\gamma = \gamma(\delta, t, r+1)$.
\end{theorem}
\begin{IEEEproof}

Note, that due to the property (\ref{eq::linearized}) the checks added by the code $\cC_E$ are evaluation points of $f(x)$ in the points of $\F_{q^m}$, that linearly depend on $\alpha_1, \alpha_2 \ldots, \alpha_{n_G}$. To decode the code $\mathcal{C}_G$ we need to interpolate $f(x)$. To do this it is sufficient to find $k$ evaluation points which correspond to linearly independent elements of $\F_{q^m}$. 

Let the code $\cC_G \diamondsuit \cC_E$ has the minimum distance $d = \delta n$, thus this code can correct any $d-1$ erasures. Let us denote the set of erasures by $E$, $|E| = d-1$, and estimate the number of evaluation points ($k'$), which correspond to linear independent elements of $\F_{q^m}$. The code $\cC_E$ imposes $\frac{t n}{r+1}$ linear restrictions. We cannot take all the evaluation points that belong to the same linear restriction as they are linearly dependent. By $\tilde{\mathbf{H}}_E$ of size $|\Gamma(E)| \times |E|$ we denote a submatrix of $\mathbf{H}_E$, which corresponds to erased positions (we removed zero rows). The probability for this submatrix to have full rank tends to $1$ when $q$ grows (see \cite{TamoBargFrolov}). Thus, the number of evaluation points corresponding to linear independent elements of $\F_{q^m}$ can be estimated as follows

\[
k' \geq n - |E| - \left(m - \min\left\{ |\Gamma(E)|, |E| \right\} \right), 
\]
where $\Gamma(E) \subset C$ is the set of linear restrictions connected to the set of erased nodes $E \subset V$. To conclude the proof we note, that $|\Gamma(E)| \geq t\gamma |E|$ and choose $k=k'$.  
\end{IEEEproof} 

\section{Concatenated construction}

\begin{figure}[t]
\centering
\includegraphics[width=0.4\textwidth]{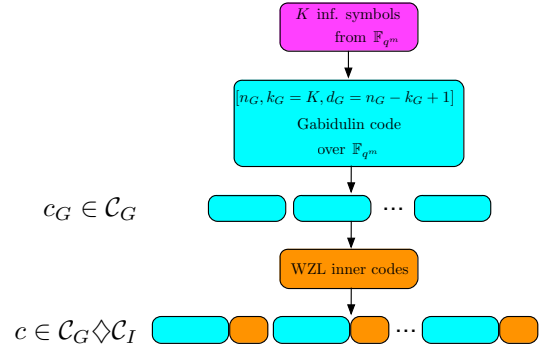}
\caption{Concatenated construction}
\label{fig::concatenated}
\end{figure}

We encode information symbols in two steps. First, $k$ information symbols over $\F_{2^m}$ are encoded using a Gabidulin code. The codeword of the Gabidulin code of length $n_G$ is then partitioned into local groups and each local group is then encoded using an $[n_I, k_I]$ binary WZL code. In what follows we assume, that $k_I | n_G$. This process is illustarted in \Fig{fig::concatenated}.

Let us consider a WZL code with parity-check matrix $\mathbf{H}_I$.  Let us denote the erasure pattern by $E_I$, $|E_I| = e_I$ and estimate the rank of the submatrix $\tilde{\mathbf{H}}_I$ of $\mathbf{H}_I$, which corresponds to erased positions. The following estimate holds 
\begin{eqnarray*}
\mathop{rank}(\tilde{\mathbf{H}}_I) &\geq& \\
L_*(e_I) &=& \left\{ \begin{gathered}
  e_I, \:\: e_I \leq t \\
  \max\left\{ \ceil*{(1-R^*(r-1,t)) e_I}, t \right\}, \:\: e_I > t \\ 
\end{gathered}  \right.
\end{eqnarray*}   
We use the fact, that the minimum code distance is $t+1$ and that the submatrix corresponds to $(\tilde{r}, t)$-LRC code with $\tilde{r} \leq r-1$.

\begin{theorem}
Let us consider a code $\cC_G \diamondsuit \cC_I$ of length $n$, minimum distance $d$ over $\F_{2^m}$. The following bound is valid for the dimension of the code 
\[
k \geq k_I \floor*{\frac{n-d+1}{n_I}} + k_I - e_I + L_*(e_I),
\]
where
\[
e_I = n_I \left( \frac{n-d+1}{n_I} - \floor*{\frac{n-d+1}{n_I}} \right).
\]
\end{theorem}
\begin{IEEEproof}
As the distance is $d$ we need to find the worst combination of $d-1$ erasures to estimate the number of evaluation points which correspond to linearly independent elements of $\F_{2^m}$. Due to the properties of WZL codes the worst combination of errors should cover the whole blocks (codewords) of inner code $\cC_I$. The number of blocks, that do not contain the erasures in this case is equal to $\floor*{\frac{n-d+1}{n_I}}$ and we can take information symbols of these blocks. In case $n_I$ does not divide $n-d+1$ we have one block which is partially erased. We can take $k_I - e_I + L_*(e_I)$ symbols from it, where $e_I$ is the number of erasures in this block.
\end{IEEEproof}

\begin{corollary}
The asymptotic from of this bound is as follows
\[
R \geq \frac{r}{r+t} (1-\delta) - o(1).
\]
\end{corollary}

\section{Numerical results}

\begin{figure}[h]
 
\begin{subfigure}{0.5\textwidth}
\center
\includegraphics[width=0.87\linewidth]{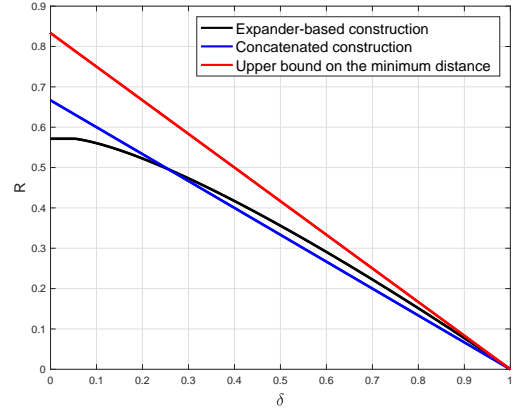} 
\caption{$t=3$, $r=6$}
\label{3-12}
\end{subfigure}
\begin{subfigure}{0.5\textwidth}
\center
\includegraphics[width=0.87\linewidth]{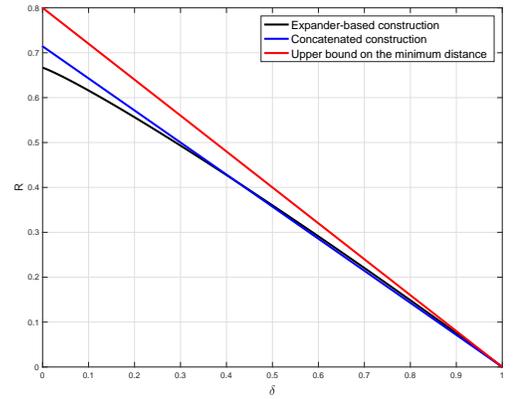}
\caption{$t=2$, $r=5$}
\label{2-5}
\end{subfigure}
\caption{Comparison of upper and lower bounds}
\label{graph}
\end{figure}

Comparison of upper and lower bounds for different values of locality and availability is presented in \Fig{graph}. 
We note, that obtained upper bound improves an upper bounds from \cite{TamoBargFrolov, Yaakobi}. Another interesting fact is as follows. In high rate region concatenated construction is better, than expander-based construction. The situation is opposite in the low rate region.

\bibliographystyle{IEEEtran}
\bibliography{main}

\end{document}